\documentclass{sig-alt-full}
\usepackage{graphicx}
\usepackage{wrapfig}

\long\def\comment#1{}
\newcommand{\commentout}[1]{}

\newcommand{\secref}[1]{Section~\ref{#1}}
\newcommand{\figref}[1]{Figure~\ref{#1}}

\begin{document}
\conferenceinfo{WOSN'08,} {August 18, 2008, Seattle, Washington, USA.}

\CopyrightYear{2008}

\crdata{978-1-60558-182-8/08/08}

\title{\LARGE \bf Analysis of Social Voting Patterns on Digg}
\author{Kristina Lerman and Aram Galstyan\\
University of Southern California \\
Information Sciences Institute\\
4676 Admiralty Way\\
Marina del Rey, California 90292, USA\\
\{lerman,galstyan\}@isi.edu }

\maketitle %\pagestyle{empty} \thispagestyle{empty}

\begin{abstract}
The social Web is transforming the way information is created and distributed. Blog authoring tools enable users to publish content, while sites such as Digg and Del.icio.us are used to distribute content to a wider audience. With content fast becoming a commodity, interest in using social networks to promote and find content has grown, both on the side of content producers (viral marketing) and consumers (recommendation).
Here we study the role of social networks in promoting content on Digg, a social news aggregator that allows users to submit links to and vote on news stories. Digg's goal is to feature the most interesting stories on its front page, and it aggregates opinions of its many users to identify them.
Like other social networking sites, Digg allows users to designate other users as ``friends'' and see what stories they found interesting. We studied the spread of interest in news stories submitted to Digg in June 2006.
Our results suggest that pattern of the spread of interest in a story on the network is indicative of how popular the story will become.
Stories that spread mainly outside of the submitter's neighborhood go on to be very popular, while stories that spread mainly through submitter's social neighborhood prove not to be very popular.
This effect is visible already in the early stages of voting, and one can make a prediction about the potential audience of a story simply by analyzing where the initial votes come from.
\end{abstract}

%\category{H.1.2}{MODELS AND PRINCIPLES}{User/Machine Systems}


% primary and secondary subject areas

\section{Introduction}
The social Web, a label that includes both social networking sites such as MySpace and Facebook and the social media sites such as Digg and Flickr, is changing the way content is created and distributed.
Web-based authoring tools enable users to rapidly publish content, from stories and opinion pieces on weblogs, to photographs and videos on Flickr and YouTube, to advice on Yahoo Answers, to Web discoveries on Del.icio.us and Furl. User-generated content is fueling the rapid expansion of the Web, accounting for much of the new Web content. In addition to allowing users to share content, social Web sites also include a social networking component, which  means that they allow users to mark other users as friends or contacts, and provide an interface to track their friends' activities, e.g., the new content they created.

With the commodification of content, content producers face a challenge of how to effectively promote and distribute their content. The challenge facing content consumers is how to efficiently identify interesting or relevant content in a vast stream of new user-generated content.
Social scientists (and marketers) have long recognized the central role social networks play in the spread of information~\cite{Granovetter}. Before the advent of electronic communication, `word--of--mouth' recommendation was carried out mainly through telephone and letter communications, or personal interactions. Modern communications technologies have further elevated the role of social networks in product recommendation~\cite{Domingos01,Richardson2002}, information dissemination~\cite{Wu03,Gruhl04} and search~\cite{mislove2006,Lerman07flickr}.
%Researchers have also noted the similarity between the spread of information on networks and epidemics. This formulation allows researchers to draw on an extensive literature studying statistical properties of the dynamics of epidemics on networks~\cite{Pastor01,Newman02}.
However, there are few empirical studies of information propagation on social networks. Even more interestingly, existing studies have produced conflicting results.
One study of music recommendation conducted in a laboratory setting, found that users' choice of music to listen to was significantly influenced by choices made by their peers~\cite{Salganik06}. However, a large-scale study of viral marketing on Amazon~\cite{Leskovec06} showed word of mouth recommendations to be largely ineffective in leading to new purchases of products. Like the previous study of information propagation through email~\cite{Wu03}, it found that most recommendation chains terminate after just a few steps. The study did note the sensitivity of recommendation to price and category of product,  leaving open the question whether social networks are an effective tool for disseminating information about, and helping users discriminate between, free (or similarly priced) products, e.g., helping users decide what blogs to read or movies to see.
%For example, is social recommendation effective in promoting product (e.g., movie, audiobook) in a subscription-based service like Netflix or Audible? Can does it affect the choices user makes about what (free) content to read online, such as news stories and blogs?

We study the role of social networks in spreading news stories on the social news aggregator Digg\footnote{http://digg.com}, which allows users to post links to and vote on news stories. Digg became popular in part because, rather than relying on an editorial board, it aggregates opinions of its many users to identify the most interesting stories online. Like other social Web sites, Digg also allows its users to create social networks by designating other users as friends, and makes it easy to track friends' activities. Through the \emph{Friends Interface}, which acts as a social recommendation engine, a user can see the stories his friends found interesting.

We perform an empirical study of social recommendation on Digg, by examining the impact of social networks on voting. 

When a story is submitted, some votes come from the network neighbors of the submitter. New votes might also attract additional votes from the neighbors of the voters, and so on. This process is analogous to a diffusion, or spread of, activation on a network.
Our results suggest that the pattern of the spread of activation through the network is indicative of how interesting the story is, or how popular it will become. If in the initial stages of voting, most  of the votes  come from within the social neighborhood of the submitter, then the story will likely not prove popular with the general Digg audience. If, on the other hand, the votes come from users not directly linked to the submitter, the story will likely prove popular.
In other words, stories which propagate mostly through the network effect,  do not carry sufficient interest for the users outside the submitter's community and will not become popular, while stories that mainly spread outside of the submitter's community will end up becoming popular. This effect is visible already in the early stages of voting, and one can make a prediction about the potential audience of a story simply by analyzing where the initial votes come from.
%We build  a simple regression tree classifier that takes only two inputs, the number of incoming links of the submitter and the number of in--network votes among the first ten, and show that it provides a reasonably good prediction accuracy.

In the section below we describe Digg, its functionality and the data we collect from it. In \secref{sec:dynamics} we empirically study the spread of interest in a story through the social network of Digg users. In \secref{sec:interestingness} we show that we can use the early stages of this spread to predict how popular the story will become.

\section{Related work}
Our findings are in line with conclusions of previous studies that showed that social networks play an important role in promoting and locating content~\cite{Lerman07flickr,mislove2007,Lerman07ic}.
%implications for Digg
In particular, Lerman~\cite{Lerman07ic} showed that users with larger social networks are more successful in getting their stories promoted to Digg's front page, even if the stories are not very interesting. These findings have implications for the design of social media and social networking sites. For example, some implementations of social recommendation may lead to the ``tyranny of the minority,'' where a small group of active, well-connected users dominate the site~\cite{Lerman07sma}. Rather than being a liability, social networks can be used to, for example, more accurately assess the quality of content, as this paper shows.

Other researchers have used Digg's trove of empirical data to study dynamics of voting. Wu and Huberman~\cite{Wu07}  found that interest in a story peaks when the story first hits the front page, and then decays with time, with a half-life of about a day. Their study is complementary to ours, as they studied dynamics of stories \emph{after} they hit the front page. Also, they do not identify a mechanism for the spread of interest in a story. We, on the other hand, propose, and empirically study, social networks as a mechanism for the spread of interest in a story. Crane and Sornette~\cite{CraneSIPS} analyzed a large number of videos posted on YouTube. By looking at the dynamics of the number of votes received by the videos, they found that they could identify high quality videos, whether they were selected by YouTube editors, or spontaneously became popular. Like Wu and Huberman, they looked at aggregate statistics, not the microscopic dynamics of the spread of interest in stories.

\section{Digg's functionality} \label{sec:digg}

The social news aggregator Digg relies on users to submit
and moderate news stories. Each new story goes to the upcoming stories queue. The new submissions (there are 1-2 new submissions every minute) are displayed in reverse chronological order, 15 to the page, with the most recent story at the top. Each day Digg selects a handful of stories to feature on its \emph{front page}.
Digg's goal is to promote only the most interesting of the submitted stories, and it relies on the opinions of its users to identify them. Digg's automatic  promotion algorithm looks at the voting patterns made within 24 hours of a story's submission. Although its details are kept secret and change on a regular basis~\cite{diggblog}, the promotion algorithm takes into account the number of votes a story receives and the rate at which it receives them, among other factors.
In the data we collected, we did not see any front-page stories with fewer than 43 votes, nor did we see any stories in the upcoming queue with more than 42 votes.

Digg also allows users to designate others as friends and makes it easy to track friends' activities. The friendship relationship is asymmetric. When user $A$ lists user
$B$ as a \emph{friend}, user $A$ is able to watch the activity of
$B$ but not vice versa. We call $A$ the \emph{fan} of $B$.
Digg provides a Friends Interface, which summarizes a user's friends' recent activity: the number of stories his friends have submitted, commented on or voted on in the preceding 48 hours. Tracking activities of friends is a feature of many social Web sites and is one of their major draws.

Digg users vary widely in their activity levels. Some users casually browse the front page, voting on one or two stories. Others spend hours a day combing the Web for new stories to submit, and voting on stories they found on Digg. Digg calculated a users' reputation based on how successful they were in getting their stories promoted to the front page. Until February 2007~\cite{roseTopUsers}, in order to encourage activity, Digg publicized users' reputation on the \textit{Top Users} list. A look at the statistics of user activity showed that top-ranked users were disproportionately active: of the more than 15,000 front page stories submitted by the top 1000 Digg users as of June 2006, the top $3\%$ of the users were responsible for $35\%$ of the submissions and a similarly high fractions of the votes cast and comments made.

\subsection{Digg dataset}
For our study, we scraped Digg's Technology section with the aid of a tool provided by Fetch Technologies. On June 30, 2006, we scraped Digg's front page, collecting data about roughly 200 of the most recently promoted stories. For each story, we extracted the story's title, name of the submitter, time the story was submitted, as well as names of users who voted on the story. Although we do not have the time stamp of each vote, they are listed in chronological order, with submitter's name appearing first on the list. In February 2008 we augmented this data with the final vote count (number of diggs) the stories received. In all, we collected information about votes from over $16,600$ distinct users.

\begin{figure}[tbh]
\begin{center}
  \includegraphics[height= 2.1in]{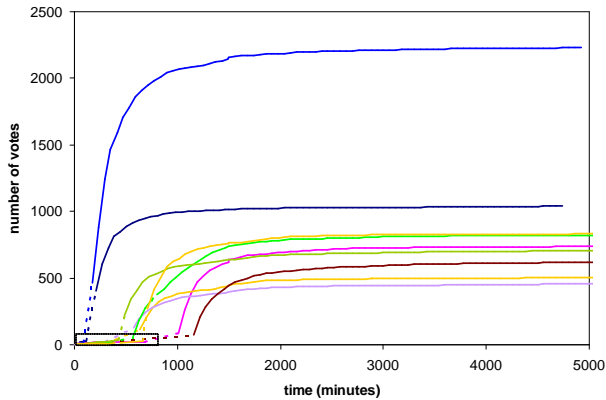}
\caption{Time series of the number of votes, since submission, received by randomly chosen front-page stories. The rectangle indicates votes received while the stories were in the upcoming stories queue, and dashes indicate transition to the front page.
} \label{fig:diggs-ts}
\end{center}
\end{figure}

\begin{figure*}[tbh]
\begin{center}
\begin{tabular}{cc}
  \includegraphics[height= 2.2in]{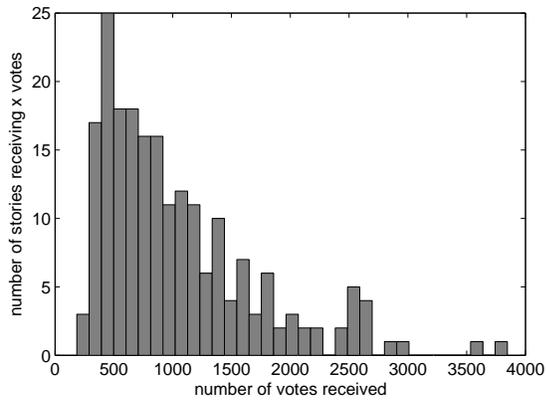} &
  \includegraphics[height= 2.2in]{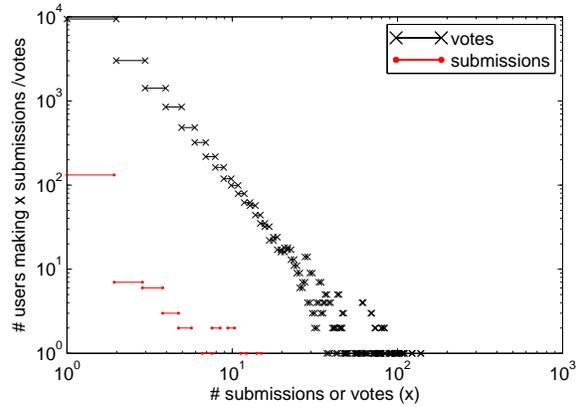}
  \\
  (a) & (b)
  \end{tabular}
\caption{Statistics of story and user activity: (a) Histogram of the number of votes received by stories. (b) Histogram of the number of stories submitted and voted on.
} \label{fig:diggs}
\end{center}
\end{figure*}

The basic dynamics of votes received by stories appears the same from story to story, as shown in \figref{fig:diggs-ts}. While in the upcoming queue (dotted rectangle in the figure), a story accumulates votes at some slow rate, but once it is promoted to the front page (indicated by dashes), it accumulates votes at a much faster rate. As the story ages, the accumulation of new votes slows down, and after a few days, the story's vote count saturates at some value. This value depends on how \emph{interesting} the story is to the general Digg community.  Some stories are very interesting, accumulating thousands votes, while others are not so interesting, receiving fewer than 500 votes.

\figref{fig:diggs}(a) shows the histogram of the final vote counts received by the front-page stories in our sample. Twenty percent of the stories were not very interesting, receiving fewer than about 500 votes, and twenty percent were very interesting, receiving more than 1500 votes. This graph is very similar to one presented by Wu and Huberman~\cite{Wu07} which showed votes received by almost 30,000 front-page stories on Digg submitted over a period of a year. In that dataset, $\sim 20\%$ of front-page stories received fewer than 400 votes, and another $\sim 20\%$ received between 400 and 600 votes. About $30\%$ of the stories received more than 1,000 votes.

The distribution of user activity is skewed, as shown in \figref{fig:diggs}(b). While most of the users had one story promoted to the front page, a number of users were responsible for multiple submissions. These were also the users with highest reputation, the so-called \emph{top users}. Voting statistics are even more skewed. While most of the users voted on only one story, some voted on many, and a few on well over a hundred stories.

\comment{
\begin{figure}[tbh]
\begin{center}
  \includegraphics[height= 2.2in]{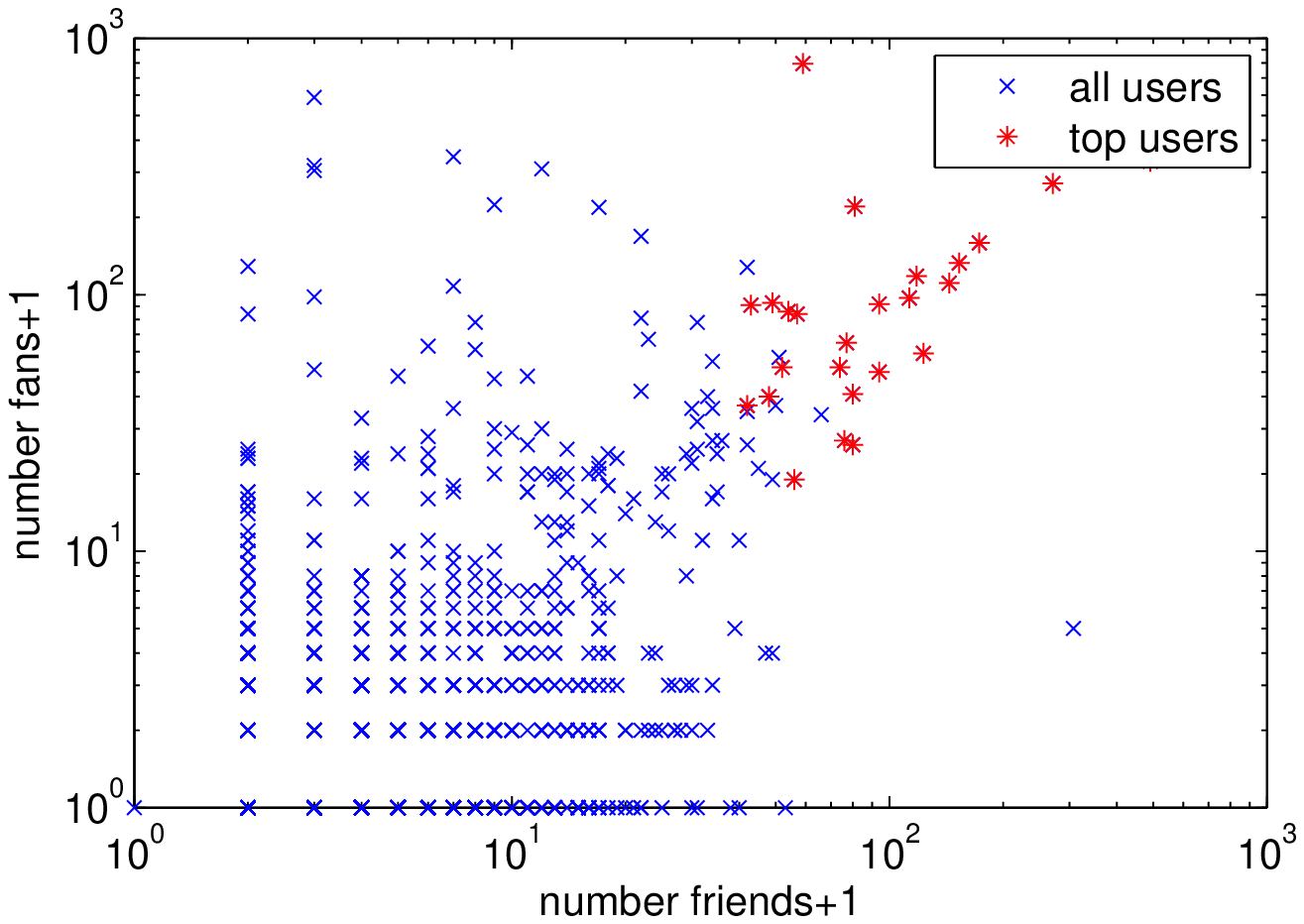}
\caption{Scatter plot of the number of friends and fans of the voters in our dataset.
} \label{fig:user-stats}
\end{center}
\end{figure}
}

\subsection{Social networks}

In addition to data about stories, we also extracted a snapshot of the social network of the top-ranked 1020 Digg users as of June 30, 2006. This data contained the names of each user's friends and fans. As a reminder, user $A$'s friends are all the users that $A$ is  watching (outgoing links on the social network graph), while $A$'s fans  are all the users watching his activity (incoming links).  Since the original social network did not contain information about all the voters in our dataset, we augmented this data in February 2008 by extracting names of fans of the $15,000+$ additional users. Many of these users acquired new fans between June 2006 and February 2008. Although Digg does not provide information about the time a fan link was created, it does list these links in reverse chronological order, with the most recent appearing on top. In addition to a fan's name, Digg also gives the date the fan joined Digg. But eliminating fans who joined Digg after June 30, 2006, we believe we were able to faithfully reconstruct the fan links (incoming edges) for all the users in our dataset.
The top users, those with most stories on the front page, tended to have more friends and fans than other users.
%\figref{fig:user-stats} shows the scatter plot of the number of friends vs fans each user has. The number was offset by one in order to be plotted on the log-log plot. The red stars correspond to 25 highest ranked users in our dataset. As can be seen from this plot, these users have greater numbers of friends and fans than other users.

\section{Information spread in \\networks}
\label{sec:dynamics}
The Digg dataset allows us to empirically study the role of social networks in the spread of information. Before a story reaches the front page, it is visible only on the upcoming stories queue and through the Friends interface. Although some users browse the upcoming stories queue, the quantity of submissions there (more than 1500 daily at the time we collected data) makes browsing unmanageable to most users. Digg also offers a visual interface to browse the upcoming and front page stories, Swarm and Stack. These visualizations are supposed to make it easier for users to identify more popular stories, but it is not clear how many users take advantage of them. Increasingly, many news sites and blogs are including a ``Digg it'' button to allow its readers to submit or vote on the story directly from the story's Web page. Again, it is not clear how many users take advantage of this option. We believe that social networks play an important role in promoting stories on Digg. In a previous work we presented data to support the claim that users employ the Friends interface to filter the vast stream of new submissions to see the stories their friends liked. Below we study the microscopics of information spread.

\begin{figure*}[tbh]
\begin{center}
\begin{tabular}{cc}
  \includegraphics[height=2.35in]{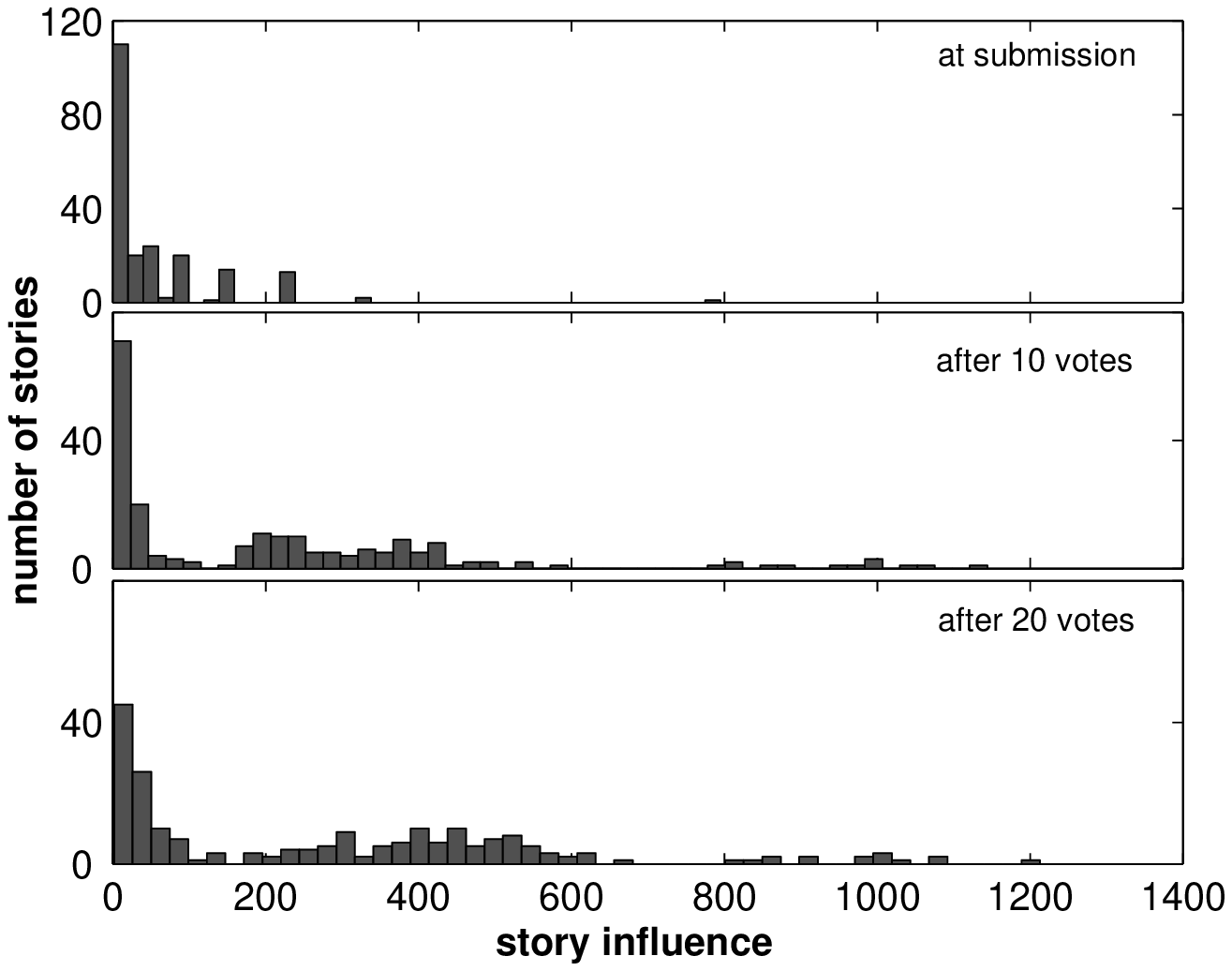} &
  \includegraphics[height=2.35in]{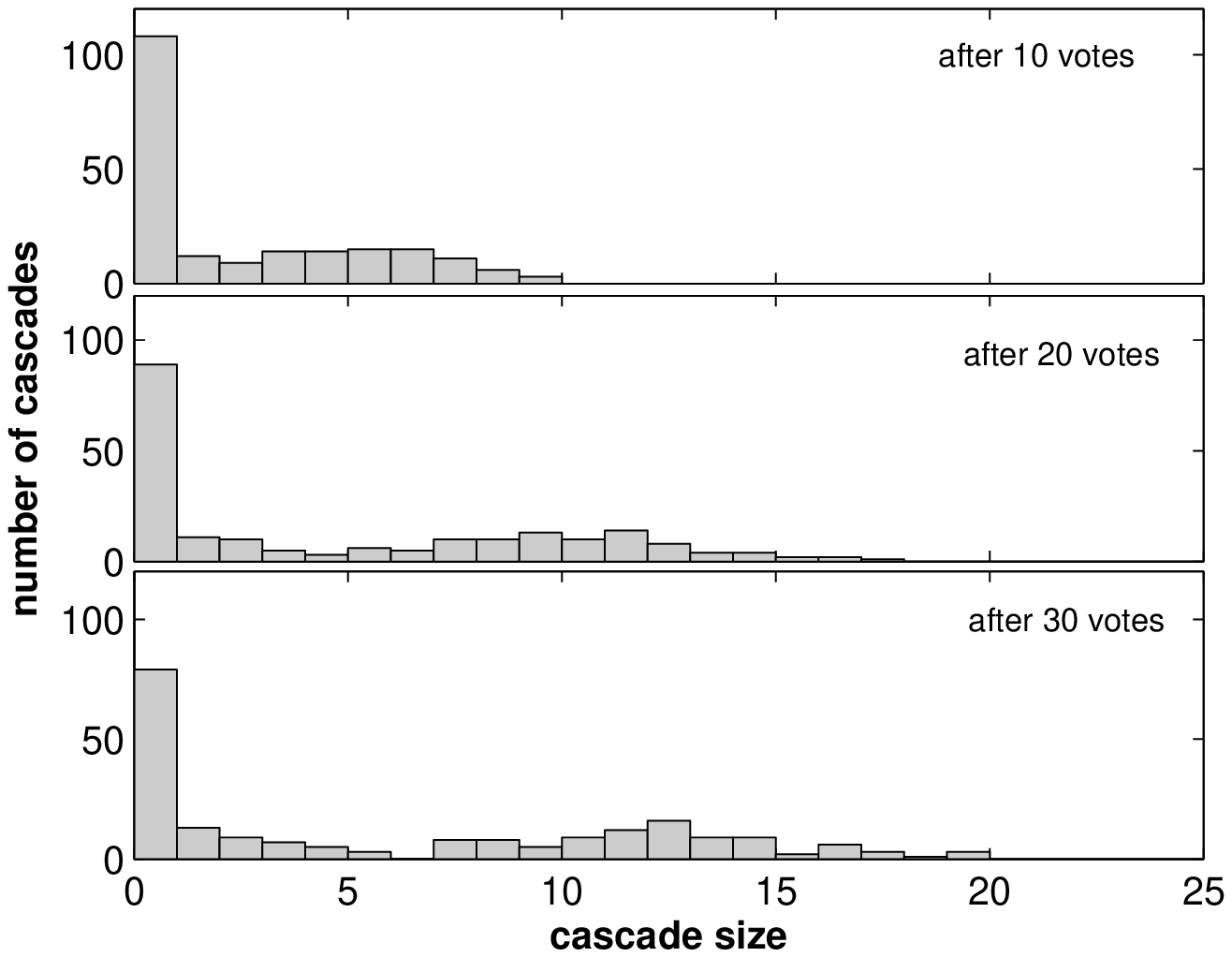}\\
  (a) & (b)
  \end{tabular}
\caption{Spread of interest in stories: (a) histogram of the story's influence, defined as the number of users who can see it through the friends interface, after it received ten votes, and (b) the number of in-network votes the story received within the first ten votes.
} \label{fig:cascades}
\end{center}
\end{figure*}

\subsection{Information cascades}
\label{sec:cascades}
At the time of submission, the story is visible only to submitter's fans through the `\emph{see the stories your friends submitted}' part of the Friends interface. As the story receives new votes, it becomes visible to many more users through the `\emph{see the stories my friends dugg}'\footnote{In this paper, as on Digg, 'digg' is synonymous with 'vote.'} part of the Friends interface. A story's \emph{influence} is given by the number of users who can see it through the Friends interface. \figref{fig:cascades}(a) shows a histogram of the stories' influence. Slightly more than half of the stories in our sample were submitted by poorly connected users with fewer than ten fans.   After stories received ten new votes, almost half of them were visible to at least 200 users through the Friends interface. After 30 votes, all the stories in our sample were visible to at least ten other users through the Friends interface, and majority of the stories were visible to hundreds of users.

Because we know the social network of Digg users, we can count how many votes came from within the network --- from fans of the previous voters. This is the story's \emph{cascade}. \figref{fig:cascades}(b) shows the distribution of cascades in our sample. For $30\%$ of the stories, at least half of the first 10 votes were in-network votes. Cascades grow with the number of votes cast. After 20 votes, $28\%$ of the stories had at least 10 in-network votes and after 30 votes, $36\%$ of the stories had at least 10 in-network votes.

\section{Story interestingness}
\label{sec:interestingness}
The total number of votes a story receives gives a measure of how \emph{interesting} it is to Digg's audience. Digg attempts to predict whether the story will be found interesting by its audience when it makes a decision whether to promote a story to the front page. It uses a number of features in the prediction, such as the number of votes received and the rate at which it receives them, Digg attempts to predict, and generally makes the prediction after 40 or so votes were cast. It is especially challenging to Digg to predict how interesting a story submitted by one of the top users is. Top users are far more active and well connected than other users, meaning that they submit and vote on many more stories, some of which happen to be stories submitted by their friends. Since top users are more likely to be in the same network, their stories are more likely to get more votes and therefore, be promoted to the front page. In September 2006, a controversy about top user dominance~\cite{techcrunch,wired} caused Digg to modify the promotion algorithm to take into account ``unique digging diversity of the individuals digging the story''~\protect\cite{diggblog}. Although this modification did result in changes in front page composition, it is not clear whether it affected the spread of interest in stories on the social networks on Digg. Rather than discounting the votes coming from fans, as Digg has chosen to do, we show that we can predict how interesting a story will be by monitoring its spread through the social network.

\subsection{Social networks and interestingness}
In a previous work~\cite{Lerman07ic} we showed that top Digg users were very successful in getting their stories promoted to the front page. We claimed that this could be explained by social browsing, i.e., the fact that Digg users use the Friends interface to find new interesting stories. We showed that social browsing, together with the observation that top users have more fans than other users, explains how less interesting stories submitted by top users are promoted to the front page. Here we study in detail how the spread of interest in a story through the social network relates to how interesting the story is.

\begin{figure}[tbh]
\begin{center}
  \includegraphics[height=3in]{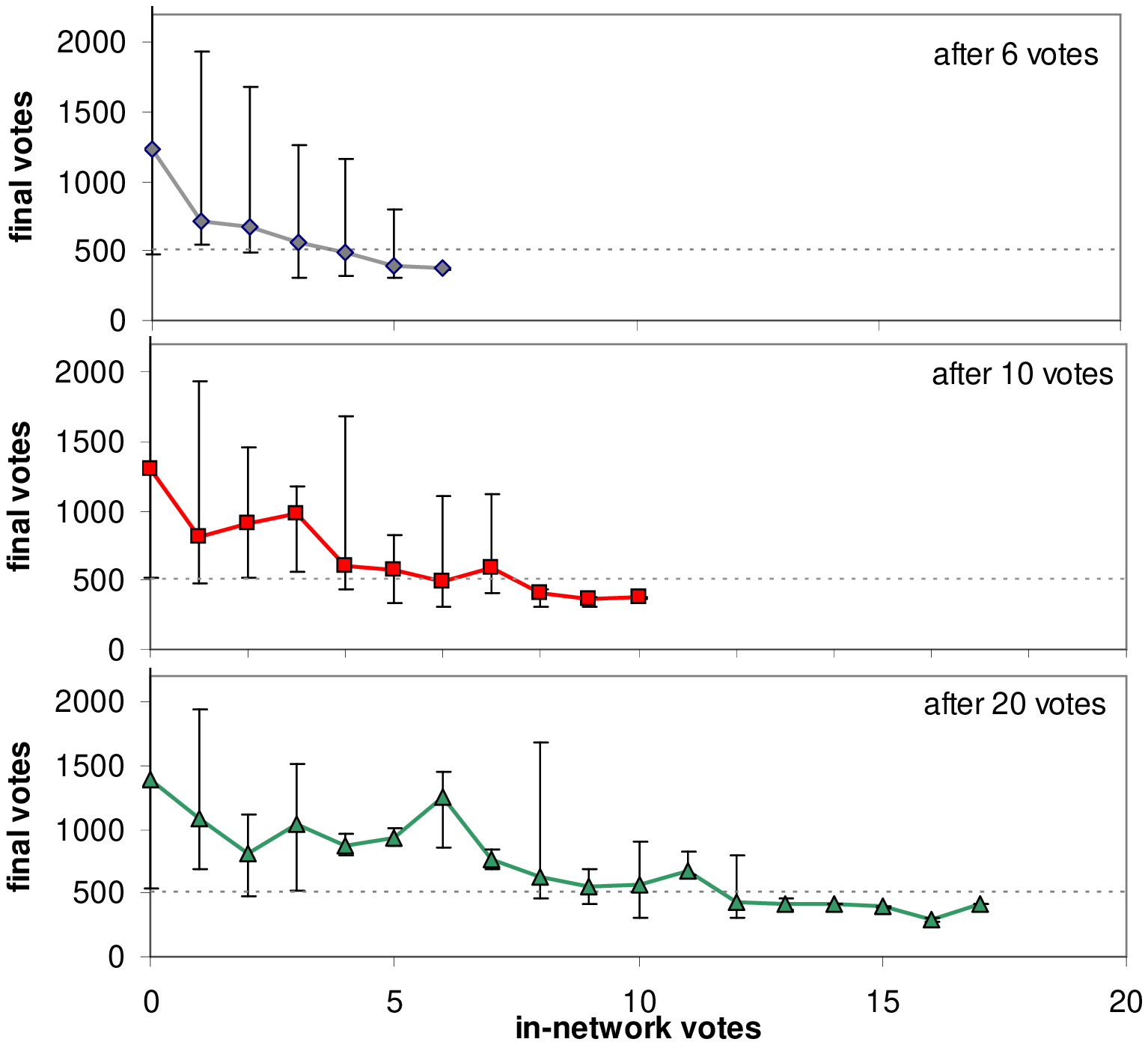}
\caption{Distribution of the number of in-network votes stories receive vs how interesting they are. The plots show the number of in-network votes received within the first (not counting the submitter) six, 10 and 20 votes.
} \label{fig:interestingness}
\end{center}
\end{figure}

\figref{fig:interestingness} shows the total number of votes a story receives (interestingness) as a function of how many in-network votes it received within the first (not counting the submitter) six, 10 and 20 votes. The graph shows the median and width of the distribution of votes  (except for the highest and lowest values) for that value of in-network votes. There is a clear inverse relationship between interestingness and the fraction of in-network votes, and this relationship is already visible early on, within the first 6--10 votes. We define a story to be interesting if it receives at least 520 votes, and not interesting if it received fewer than 520 votes.\footnote{This threshold was chosen based on \protect\figref{fig:diggs}(a), which indicated that 20\% of the stories in the sample received fewer than 500 votes, suggesting 500 as the interestingness threshold. Two stories in our sample that were submitted by top users were close to this threshold, with 505 and 507 votes (with five in-network votes each). We made the decision to raise the interestingness threshold to 520 and keep these ambiguous cases in the sample.}
As found in our previous work, many of the front-page stories submitted by best connected (top) users were deemed to be uninteresting, receiving fewer votes, and almost all of the stories submitted by poorly connected users were found to be highly interesting, with many gathering thousands of votes. One of the exceptions was a story submitted by a poorly connected user that gathered only 185 votes. One of the early voters for this story was \emph{kevinrose}, the founder of Digg and the user with most fans. The extra visibility that \emph{kevinrose}'s vote gave to the story, helped promoted this uninteresting story to the front page.

These observations suggest that there are two mechanisms for the spread of interest in a story on Digg: interest-based and  network-based. A highly interesting story will spread from many independent seed sites, as users unconnected to network of the previous voters discover the story with some small probability and propagate it to their own fans. A story that is interesting to a narrow community, however, will spread within that community only, without being picked up by unconnected users.

\subsection{Predicting interestingness}
\label{sec:prediction}
The evidence presented in the section above suggests that it is possible to predict how interesting a story is by monitoring how interest in it spreads through the social network. Moreover, it should be possible to make the prediction relatively early, after the first 6--10 votes. Digg generally waits longer, until a story accumulates at least 40 votes. Such prediction is especially useful for stories submitted by top users who tend to have bigger and more active social networks, and therefore, make it more difficult to decipher between a user's popularity and story interestingness.

\begin{figure}[tbh]
\begin{center}
  \includegraphics[height=1.8in]{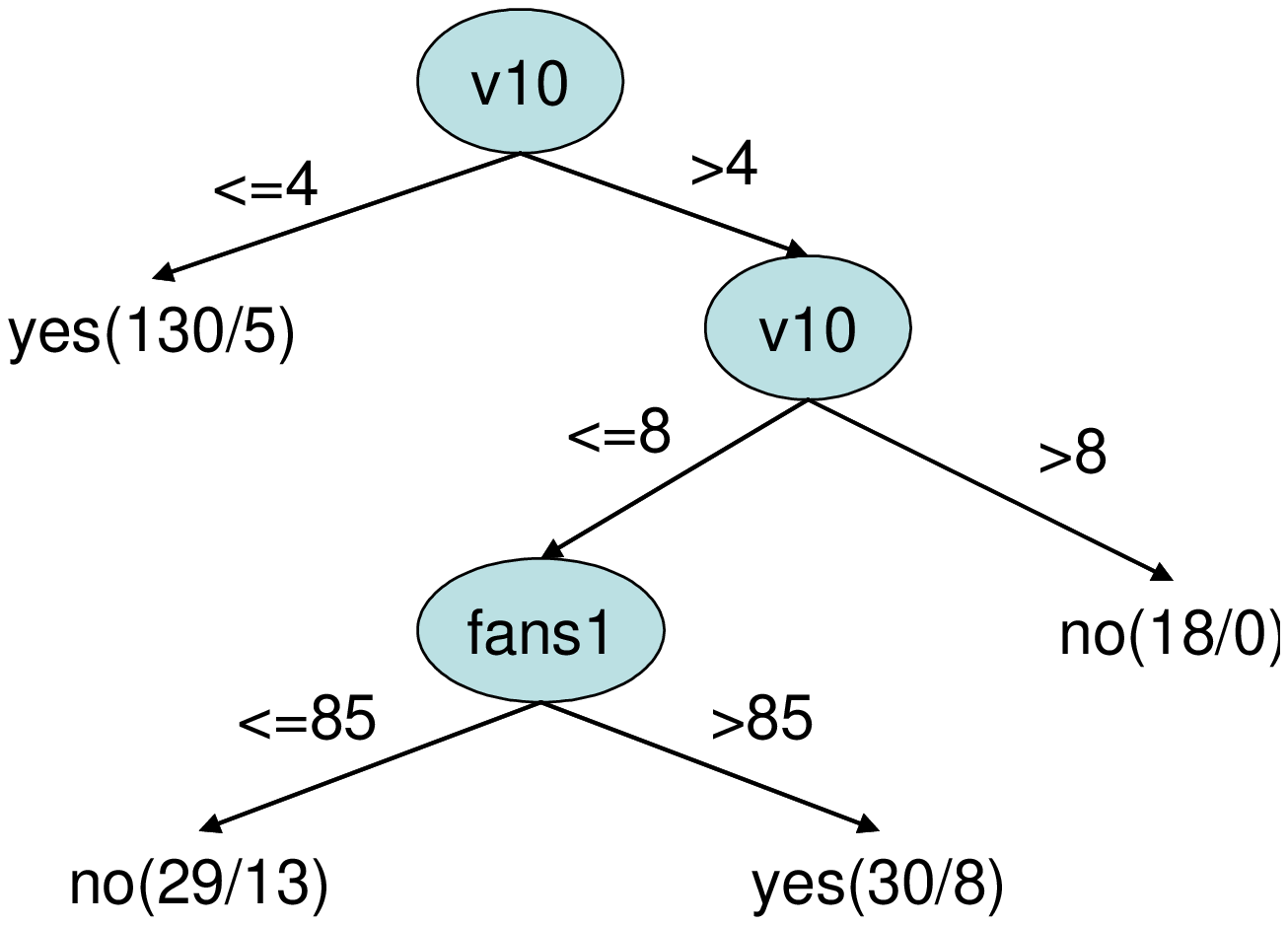}
\caption{Decision tree classifier trained on the votes data.
} \label{fig:tree}
\end{center}
\end{figure}

We trained a C4.5 (J48) decision tree classifier~\cite{Weka} on 207 stories in our dataset. Each story had three attributes: number of in-network votes within the first ten votes (v10), number of users watching the submitter (fans1) and a boolean attribute indicating whether the story was interesting or not. The story was judged interesting if it received more than 520 votes. \figref{fig:tree} shows the learned decision tree. Results of 10-fold validation indicate that this tree correctly classifies 174 of the examples, and misclassifies 33 examples.

We tested the learned model on stories extracted from the upcoming queue on June 30, 2006. This dataset consisted of 900 stories submitted within the same time period as the data analyzed above, but not yet promoted to the front page. We augmented this data by retrieving the final number of votes received by stories. From this set, we kept only the stories that were submitted by top users (with rank $\le 100$) and received at least 10 votes, leaving 48 stories.

We used the learned classifier in \figref{fig:tree} to predict whether a story was interesting (received more than 520 votes). The classifier correctly predicted 36 examples (TP=4, TN=32) and made 12 errors (FP=11, FN=1).\footnote{The notation denotes true positives (TP), true negatives (TN), false positives (FP) and false negatives (FN).} It is difficult to compare the predictions made by our  algorithm to those made by Digg, because some of the stories that Digg did not promote could have ended up receiving many votes and being deemed interesting. When we limit the comparison only to the stories that Digg did promote, of the 14 stories promoted by Digg, only five went on to receive more than 520 votes (P=TP/(TP+FP)=0.36), in other words, were judged as interesting by the Digg community. In contrast, our algorithm said that seven of these stories were interesting, and of these four received more than 520 votes (P=0.57).

\section{Conclusion}
We studied empirically the spread of interest in news stories on the social news aggregator Digg. We found that social networks play a significant role in promoting stories. In addition, we show that the pattern of social voting can be used for predicting how interesting the story will be.
Although our study was carried out on data from Digg, we believe that its conclusions will apply to other social media sites that use social networks to promote content.

As a future work, it will be interesting to  analyze more thoroughly the role of network's structural properties on the voting dynamics. Indeed, it is known that structural properties can have a significant impact on  various dynamical processes on networks.  For instance,  it is known that power--law degree distribution observed in many real--world networks can lead to  vanishing threshold for epidemics~\cite{vespignani2001a,vespignani2001b} for certain models, in a sharp contrast with the results for random Erdos-Renyi networks. Furthermore, the presence of well--connected clusters of nodes  can impact the transient dynamics of various influence propagation models\cite{galstyan2007_PRE}. This latter phenomenon can be especially important in networks with well--defined {\em community structure}~\cite{girvan2002,newman2006}.

%\bibliographystyle{plain}
%\bibliography{diggflow}

\end{document}